\documentclass[onecolumn]{sn-jnl}
\usepackage{geometry}
\usepackage{graphicx}%
\usepackage{multirow}%
\usepackage{amsmath,amssymb,amsfonts}%
\usepackage{amsthm}%
\usepackage{booktabs}%
\usepackage{mathrsfs}%
\usepackage[title]{appendix}%
\usepackage{xcolor}%
\usepackage{subcaption}
\usepackage{textcomp}%
\usepackage{manyfoot}%
\usepackage{placeins}
\usepackage{caption}%
\usepackage{algorithm}%
\usepackage{algorithmicx}%
\usepackage{algpseudocode}%
\usepackage{listings}%

\theoremstyle{plain}
\newtheorem{theorem}{Theorem}[section]

\begin{document}

\title{Regular Black Holes from Collapsing Dust in a Dark Energy Background}


\author[1,2]{\fnm{Qazi Abdul} \sur{Ghafoor}}\email{qaghafoor08@mail.ustc.edu.cn}


\affil[1]{\orgdiv{Department of Astronomy}, \orgname{University of Science and Technology of China}, \city{Hefei}, \postcode{230026}, \state{Anhui}, \country{China}}

\affil[2]{\orgdiv{School of Astronomy and Space Science}, \orgname{University of Science and Technology of China}, \city{Hefei}, \postcode{230026}, \state{Anhui}, \country{China}}

\affil[1,2]{\textbf{Corresponding Author:} Qazi Abdul Ghafoor, Email: \href{mailto:qaghafoor08@mail.ustc.edu.cn}{qaghafoor08@mail.ustc.edu.cn}}

\abstract{The central singularity in gravitational collapse represents a fundamental breakdown of classical general relativity, yet its resolution remains an open challenge. While conventional regular black hole models invoke exotic matter or nonlinear electrodynamics, the physical origin of such regularization mechanisms remains obscure. In this work, we demonstrate that a radially varying interaction between dust and dark energy can naturally resolve the central singularity. We systematically investigate three cases: no interaction, constant interaction, and radially varying interaction. Only when the interaction grows sufficiently toward the center can the divergent behavior of the density and the central mass contribution simultaneously be eliminated near the center. This yields a mass function scales as $M \sim \chi^3$, a finite Kretschmann scalar ($\lim_{\chi \to 0} K < \infty$), and a de Sitter-like core for the physical regime $w \leq -1$ (with $w=-1$ corresponding to pure de Sitter core). At the center, the resulting configuration satisfies the null, weak, and dominant energy conditions, while the strong energy condition is violated. Unlike charge-based models, our interaction parameter $\alpha(t)$ stabilizes at a non-zero value after the collapse, allowing the regular core to persist in the late-time configuration. This provides a physically motivated alternative to conventional regular black hole models, grounded in the modern cosmological understanding of dark energy rather than exotic fields or charges.}


\makeatletter
\let\headerps@out\@gobble
\makeatother

\maketitle

\section{Introduction}
The study of gravitational collapse has a long, rich history, beginning with the seminal work of Oppenheimer and Snyder in 1939 \cite{Oppenheimer_1938}, who showed that a homogeneous dust sphere collapses to form a black hole in general relativity. Thus, dust collapse is an extremely favorable case and an old problem. After that, numerous studies have been performed on dust collapse in Einstein's gravity and several modified gravities. 
In 2019, the Event Horizon Telescope collaboration captured the first image of a black hole shadow at the center of the M87 galaxy \cite{EventHorizonTelescope_2019}, and three years later, they obtained images of Sagittarius A*, the supermassive black hole at the center of our Milky Way \cite{EventHorizonTelescope_2022}. These observations have moved black holes from the class of exotic theoretical objects to the class of real astrophysical ones. However, the singularity at the center of a black hole indicates that general relativity is not a complete theory to describe processes in the vicinity of this region. Moreover, Penrose's theorem on singularities \cite{Penrose_1965} states that during gravitational collapse, when the apparent horizon forms, a singularity is also inevitably formed. \\
Recently, regular black holes — solutions in which the central singularity is replaced by a de Sitter core — have attracted considerable attention \cite{Ansoldi_2008, Lan_2023, Bonanno_2024, Bronnikov_2001, Konoplich_1999, Khlopov_2000, bueno25}. The idea that superdense matter passes into a vacuum medium belongs to Gliner \cite{Gliner_1966} and Sakharov \cite{Sakharov_1966}. Bardeen was the first who constructed the model of a regular black hole \cite{Bardeen_1968}, and it was later found that this solution was supported by nonlinear electrodynamics. After that, other important models of regular black holes were found such as Hayward black hole \cite{Hayward_2006}, Dymnikova black hole \cite{Dymnikova1992} and others. Recently, it has been found that the Hagedorn fluid can also be the source of a regular black hole \cite{Vertogradov_Ovgun_2025}. However, a black hole is formed as a result of the gravitational collapse of an ordinary massive star, inside which there is no such exotic matter. \\
In the recent decade, similar studies were performed by Cai and Wang \cite{Cai_2006} who studied the gravitational collapse of a dust cloud in a dark energy background, and their results have been extended to anisotropic fluid in Ref. \cite{shah_2018}. Many studies have investigated the collapse of dust fluid and dark energy (DE), extending general relativity to scenarios including modified gravity \cite{Cai_2006}. It was found that black holes can form due to condensation of the dust fluid, and this result remains true even when the interaction between dust and dark energy does not vanish. When \( w < -1 \) (phantoms), their models can be interpreted as representing the death of a white hole that ejects both dust and phantoms, with the ejected matter recollapsing to form a black hole. In modified gravity and general relativity, recent studies have looked into the collapse with the interacting and non-interacting combination of dust and dark energy \cite{Pandey_2022, Ahmad_2013, Chakraborty_2010}. Recently, the analysis of dust collapse in dark energy backgrounds has been extended to inhomogeneous configurations \cite{Pandey_2022}, revealing that the outcome of collapse depends sensitively on the DE equation of state and initial density profile. Specifically, when DE equation of state satisfies certain conditions, collapse leads to black hole formation, while for other values, collapse may be prevented or proceed to naked singularity \cite{Pandey_2022, Cai_2006}. A fully relativistic approach to structure formation from gravitational collapse incorporating both DE and dust has been established in recent studies \cite{Koushiki_2025}, deriving the scalar field potential required for equilibrium without depending on Newtonian virialization. \\
Black holes play an important role in structure formation in the universe, and are described by spacetime singularities enclosed within an event horizon. The presence of dark energy, with its extremely negative pressure, raises important questions about its impact on large scale structure formation and gravitational collapse. It introduces repulsive gravitational effects that can potentially influence the dynamics of massive clouds. Although such effects are generally negligible at astrophysical scales, they become significant at cosmological scales, where dark energy may inhibit collapse or even give rise to exotic compact objects, such as dark energy stars or gravastars \cite{chapline2005darkenergystars}. Understanding the interplay between gravitational collapse is therefore essential for a complete picture of structure formation and evolution of black holes. However, the interplay between dust and dark energy becomes challenging in collapse scenarios. The findings from the recent studies have revealed that dust collapse in a background of dark energy forms a black hole \cite{Cai_2006, soma2018}. The black hole development is consistent with Horava-Lifshitz gravity when DE and dust combine, as evidenced by findings showing that DE alone decreases the likelihood of black hole formation \cite{Cai_2006, rudra2014gravitational, soma2018}. Additionally, the effect of dark energy on black holes that have already formed due to radiation or dust collapse is an emerging area now. Bronnikov and Fabris have systematically studied regular phantom black holes, identifying 16 classes of possible regular configurations including asymptotically flat black holes where the singularity is replaced by a de Sitter infinity \cite{Bronnikov2006}. Babichev et al. \cite{Babichev_2005} showed that dark energy accretion can drastically minimize a black hole's mass, possibly to zero in phantom energy context. This mass decrease has been further investigated in several studies. Later, Bronnikov, Fabris and Gon\c{c}alves \cite{Bronnikov2007} presented that local concentrations of a phantom field can form regular black holes with asymptotically flat static regions, separated by an event horizon from an expanding, singularity-free, asymptotically de Sitter universe. The well-known regular black hole solution with a de Sitter core, originally proposed by Dymnikova \cite{Dymnikova1992}, discusses a limiting case \( w = -1 \) where dark energy reduces to a cosmological constant, the interaction vanishes at the center, and the core is composed purely of dark energy, yielding a nonsingular black hole. The charged dust and dark energy interactions has been shown to produce nonsingular interior solutions matching the Reissner-Nordström de Sitter exterior geometry \cite{Maier_2020}. Fabris and Pav\'on considered a cosmological scenario dominated by phantom dark energy, and black holes condensing out of this component \cite{Fabris2009}. They demonstrated that the big rip singularity can be avoided via black hole production, showing that phantom fluids can lead to regular configurations rather than singularities. Furthermore, the gravitational collapse study has been extended to higher order gravity theories such as Gauss-Bonnet gravity, where general initial conditions have been used to analyze black hole horizons \cite{EPJC_2025}. \\
Recent work on regular black hole formation from collapsing dust and radiation \cite{Vertogradov_2025} has opened new perspectives on the possible end states of gravitational collapse. However, this study was limited to radiation with a fixed equation of state ($w=1/3$). The role of dark energy, which constitutes the dominant component of the universe's energy budget, remains unexplored in this context. In this paper, we generalize the interaction mechanism to the case of dust and dark energy. We systematically investigate three cases: no interaction, constant interaction model, and radially varying interaction model. This investigation not only extends the scope of the regularization mechanism but also reveals the central role of dark energy in resolving the singularity. A key result of this work is the establishment of a regularity criterion for physical regular black holes: we show that a non-negative central dust density requires $w \leq -1$, so that physically admissible solutions exist only in the phantom regime $w < -1$ or at the cosmological constant limit $w = -1$. This criterion is formalized in Theorem~\ref{regularity}. \\
The remainder of this paper is organized as follows: Section~II describes background geometry and matter content. Section~III analyzes the non-interacting dust and dark energy model. Section~IV presents the constant interaction between dust and dark energy. Section~V examines the radially varying interaction and provides an explicit example of a regular black hole solution. Section~VI discusses the energy conditions for the explicit example of radially varying interaction model. Section~VII presents a discussion of the results, and Section~VIII summarizes the main findings and concludes the study. 
\section{Spacetime Metric and Matter Content}
We investigate the gravitational collapse of a spherically symmetric cloud consisting of two interacting components: pressureless dust and dark energy. The main purpose is to study whether an energy exchange between these two sectors can modify the central singularity structure of the collapsing configuration. \\

We describe the spacetime by the generalized Eddington-Finkelstein metric
\begin{eqnarray} \label{1}
ds^2 = -f(t,\chi)dt^2 + 2 dt\,d\chi + \chi^2 d\Omega^2,
\end{eqnarray}
where
\begin{eqnarray} \label{2}
f(t,\chi) = 1 - \frac{2M(t,\chi)}{\chi}.
\end{eqnarray}
Here $\chi$ denotes the areal radius and $M(t,\chi)$ is the generalized mass function. The use of Eddington-Finkelstein coordinates allows us to describe both the collapse phase and the formation of trapped surfaces without coordinate singularities at the horizon. The metric function $f(t,\chi)$ determines the location of apparent horizons through the condition $f(t,\chi)=0$, which corresponds to $2M(t,\chi)=\chi$.  \\

The energy density and the isotropic pressure associated with this geometry are
\begin{eqnarray} \label{3}
\rho_t = \frac{2M'}{\chi^2},
\end{eqnarray}
and
\begin{eqnarray} \label{4}
P = -\frac{M''}{\chi},
\end{eqnarray}
while the time variation of the mass function generates an energy flux
\begin{eqnarray}\label{5}
\sigma = \frac{2\dot M}{\chi^2}.
\end{eqnarray}
Here primes ($'$) denote derivatives with respect to $\chi$ and dots denote derivatives with respect to $t$. The quantity $\sigma$ represents the energy flux associated with the collapsing matter, which becomes important when the mass function depends explicitly on time.\\

The collapsing matter distribution is assumed to consist of two sectors,
\begin{eqnarray} \label{6}
\rho_t = \rho_m + \rho_{\rm de},
\end{eqnarray}
\begin{eqnarray} \label{7}
P=P_m + P_{de}
\end{eqnarray}
where $\rho_m$ represents the dust component with satisfying $P_m=0$,  and $\rho_{\rm de}$ denotes the dark-energy contribution obeying
\begin{eqnarray} \label{8}
P_{\rm de} = w \rho_{\rm de},
\end{eqnarray}
with $w < 0$ for dark energy. The conservation equation follows from the vanishing divergence of the energy-momentum tensor,
\begin{eqnarray} \label{9}
\nabla_\mu T^{\mu\nu} = 0,
\end{eqnarray}
and for the total system becomes
\begin{eqnarray} \label{10}
\rho_t' + \frac{2}{\chi}(\rho_t + P) = 0,
\end{eqnarray}
Eq.~(\ref{7}) expresses the local conservation of energy for the combined fluid. The interaction between the two sectors determines whether this conservation law can be decomposed into independent conservation equations or whether an exchange term must be introduced.

\section{Non-Interacting Dust and Dark Energy}

We first consider the simplest scenario in which the dust and dark-energy components evolve independently during the gravitational collapse. In this case, although the total energy-momentum tensor is conserved, each component separately satisfies a conservation equation. This corresponds to the absence of any energy exchange between the two sectors. \\

The separate conservation equations are therefore
\begin{eqnarray} \label{11}
\rho_m' + \frac{2}{\chi}\rho_m = 0,
\end{eqnarray}
and
\begin{eqnarray} \label{12}
\rho_{\rm de}' + \frac{2}{\chi}(\rho_{\rm de} + P_{\rm de}) = 0.
\end{eqnarray}
Using the dark-energy equation of state $P_{\rm de} = w\rho_{\rm de}$, the equation~(\ref{12}) becomes
\begin{eqnarray} \label{13}
\rho_{\rm de}' + \frac{2(1+w)}{\chi}\rho_{\rm de} = 0.
\end{eqnarray}
Solving these equations gives the radial dependence of the two components:
\begin{eqnarray} \label{14}
\rho_m(t,\chi) = \rho_m^0(t)\chi^{-2},
\end{eqnarray}
and
\begin{eqnarray} \label{15}
\rho_{\rm de}(t,\chi) = \rho_{\rm de}^0(t)\chi^{-2(1+w)}.
\end{eqnarray}
The dust density scales as $\chi^{-2}$, characteristic of the pressureless collapse considered here, while the dark-energy contribution scales as $\chi^{-2(1+w)}$. For $w=-1$, the dark-energy density becomes radially constant, corresponding to a vacuum-energy contribution. However, since the dust density still scales as $\chi^{-2}$, the total density remains singular at the center as long as the dust component is present. For $w<-1$, the exponent $-2(1+w)$ becomes positive, and the dark-energy density decreases toward the center. Thus, in the phantom regime, the dust contribution remains dominant near the center. \\

The total energy density is therefore
\begin{eqnarray} \label{16}
\rho_t(t,\chi) = \rho_m^0(t)\chi^{-2} + \rho_{\rm de}^0(t)\chi^{-2(1+w)}.
\end{eqnarray}
While the pressure is
\begin{eqnarray} \label{17}
P(t,\chi) = w \rho_{\rm de} = w \rho_{\rm de}^0(t)\chi^{-2(1+w)}.
\end{eqnarray}
Using the relation between the mass function and the total density, $\rho_t = 2M'/\chi^2$, we obtain
\begin{eqnarray} \label{18}
2M' = \rho_m^0(t) + \rho_{\rm de}^0(t)\chi^{-2w}.
\end{eqnarray}
Integrating with respect to $\chi$, the mass function becomes
\begin{eqnarray} \label{19}
M(t,\chi) = M_0(t) + \frac12\rho_m^0(t)\chi + \frac{\rho_{\rm de}^0(t)}{2(1-2w)}\chi^{1-2w},
\end{eqnarray}
where $M_0(t)$ represents the central contribution to the mass function. Consequently, the metric function (\ref{2}) takes the form
\begin{eqnarray} \label{20}
f(t,\chi) = 1 - \frac{2M_0(t)}{\chi} - \rho_m^0(t) - \frac{\rho_{\rm de}^0(t)}{1-2w}\chi^{-2w}.
\end{eqnarray}
The apparent horizon is obtained from the condition $f(t,\chi_H) = 0$, which gives
\begin{eqnarray} \label{21}
1 - \frac{2M_0(t)}{\chi_H} - \rho_m^0(t) - \frac{\rho_{\rm de}^0(t)}{1-2w}\chi_H^{-2w} = 0.
\end{eqnarray}
The above solution describes a collapse in which dust and dark energy do not exchange energy. The dust component follows the usual $\chi^{-2}$ collapse scaling, whereas the dark-energy contribution depends on the equation-of-state parameter $w$. In particular, for $w=-1$, the dark-energy density is radially constant and provides a vacuum-energy contribution. Nevertheless, the dust contribution continues to diverge as $\chi^{-2}$ toward the center, so the non-interacting solution does not produce a regular core. \\

For phantom dark energy, $w<-1$, the dark-energy density decreases toward the center, and the dust component remains the dominant contribution to the central density. More generally, for the usual dark-energy range $-1\leq w<-1/3$, the dust contribution scales as $\chi^{-2}$ and dominates the central behavior. The mass function $M$ from equation~(\ref{19}) consequently contains a term linear in $\chi$, near the center. A regular spherical center requires the mass function to behave at least as $M(t,\chi)=\mathcal{O}(\chi^3)$. The non-interacting solution therefore fails to satisfy the regularity condition for a nonzero dust density. The singularity structure can consequently be examined through the curvature invariants.
\subsection{Curvature Invariants and Singularity Structure}
Equation~(\ref{20}) is the combination of Husain like solution for dark energy and dust \cite{vitali95, hussain95, Vertogradov_2025}. A black hole and at least one apparent horizon can be described by this solution. In order to analyze whether this solution is nonsingular or not, it is important to calculate curvature invariants, the ricci scalar $R$, the square Ricci tensor $S$ and the Kretschmann scalar $K$. The curvature invariants provide important information about the strength and nature of the singularity formed during gravitational collapse. For the generalized Vaidya geometry, the Ricci scalar is expressed in terms of the total energy density and pressure by the metric (\ref{1}) as, 
\begin{eqnarray} \label{22}
R = 2(\rho_t - P),
\end{eqnarray}
\begin{eqnarray} \label{23}
R_{\mu\nu}R^{\mu\nu} = 2\rho_t^2 + 2P^2,
\end{eqnarray}
and the Kretschmann scalar takes the form
\begin{eqnarray} \label{24}
K = \frac{48M^2}{\chi^6} - \frac{16M}{\chi^3}(2\rho_t - P) + 8(\rho_t^2 - \rho_t P + \frac{1}2{}P^2).
\end{eqnarray}
Equivalently, using Eqs.~(\ref{3}) and (\ref{4}), these invariants can be written explicitly in terms of $M$ and its derivatives as
\begin{eqnarray} \label{25}
R = \frac{4M' + 2\chi M''}{\chi^2},
\end{eqnarray}
\begin{eqnarray} \label{26}
R_{\mu\nu}R^{\mu\nu} = \frac{8M'^2 + 2\chi^2 M''^2}{\chi^4},
\end{eqnarray}
and
\begin{eqnarray} \label{27}
K = \frac{8(6M^2 - 8\chi M M' + 4\chi^2 M'^2 + 2\chi^2 M M'' - 2\chi^3 M' M'' + \frac{1}{2}\chi^4 M''^2)}{\chi^6}.
\end{eqnarray}
The behavior near the center is determined by the competition between the dust and dark-energy contributions. Since the dust component scales as $\rho_m^0(t)\chi^{-2}$, it dominates over the dark-energy contribution for the usual dark-energy range $-1 \leq w < -1/3$. Consequently, the  total density (\ref{16}) near the center is
\begin{eqnarray} \label{28}
\rho_t = \rho_m^0(t)\chi^{-2} + \mathcal{O}\left(\chi^{-2(1+w)}\right).
\end{eqnarray}
The mass function (\ref{19}) behaves as
\begin{eqnarray} \label{29}
M(t,\chi) = M_0(t) + \frac12\rho_m^0(t)\chi +  \mathcal{O}\left(\chi^{1-2w)}\right).
\end{eqnarray}
If the central mass contribution is non-zero, $M_0(t) \neq 0$, the Kretschmann scalar yields
\begin{eqnarray} \label{30}
\lim_{\chi \to 0} K = \infty,
\end{eqnarray}
which demonstrates the presence of a curvature singularity at the center. \\
If the central mass contribution is set to zero, $M_0(t)=0$, then the mass function (\ref{19}) becomes
\begin{eqnarray} \label{31}
M(t,\chi) = \frac12\rho_m^0(t)\chi + \frac{\rho_{\rm de}^0(t)}{2(1-2w)}\chi^{1-2w}.
\end{eqnarray}
The Kretschmann scalar still diverges. Hence, the non-interacting configuration remains singular and does not satisfy the regular-center conditions. Physically, this result occurs because the dust and dark-energy sectors evolve independently. The pressureless component continues to collapse according to the $\chi^{-2}$ scaling without any energy exchange with the dark-energy sector that could modify its central behavior. Although dark energy contributes negative pressure, the absence of interaction does not alter the dominant dust contribution near the center. Consequently, the dust-dominated collapse drives the system toward a curvature singularity. \\

Hence, the non-interacting model provides the singular reference solution. In order to investigate whether the central singularity can be weakened or removed, we next introduce an explicit interaction between the dust and dark-energy sectors.

\section{Constant Interaction Model}

In the previous section, we studied the collapse of a cloud composed of dust and dark energy in the absence of interaction between the two components. In that case, both sectors evolve independently and the gravitational dynamics is determined by their separate conservation laws. We now consider a more general scenario in which dust and dark energy exchange energy during the collapse. The energy transfer between the two components can modify the effective pressure and the gravitational mass of the collapsing system, potentially changing the nature of the central singularity. \\

Although the individual components are not conserved separately, the total energy-momentum tensor remains conserved. Therefore, the total energy density $\rho_t = \rho_m + \rho_{\rm de}$ satisfies the conservation equation (\ref{10}). The interaction between the two sectors is introduced through a constant interaction parameter $\epsilon$. The conservation equations for the individual components become \cite{Vertogradov_2025}
\begin{eqnarray} \label{32}
\rho_m' + \frac{2}{\chi}\rho_m = -\frac{\epsilon}{\chi}\rho_{\rm de},
\end{eqnarray}
and
\begin{eqnarray} \label{33}
\rho_{\rm de}' + \frac{2(1+w)}{\chi}\rho_{\rm de} = \frac{\epsilon}{\chi}\rho_{\rm de}.
\end{eqnarray}
Here $\epsilon > 0$ describes the transfer of energy from the dust component to the dark-energy sector, while $\epsilon < 0$ corresponds to the transfer of energy in the opposite direction. Since dark energy possesses negative pressure, such an energy exchange can influence the competition between the gravitational attraction of dust and the repulsive effect of dark energy.

\subsection{Density Profiles}

The dark-energy conservation equation (\ref{33}) can be written as
\begin{eqnarray} \label{34}
\frac{\rho'_{\rm de}}{\rho_{\rm de}} = \frac{\epsilon - 2(1+w)}{\chi}.
\end{eqnarray}
Integrating with respect to $\chi$, we obtain
\begin{eqnarray} \label{35}
\ln(\rho_{\rm de}) = (\epsilon - 2(1+w))\ln(\chi) + \alpha(t),
\end{eqnarray}
where $\alpha(t)$ is a function of integration that depends only on time. Therefore, the dark-energy density becomes
\begin{eqnarray} \label{36}
\rho_{\rm de}(t,\chi) = \rho_{\rm de}^0(t)\chi^{\epsilon - 2(1+w)}.
\end{eqnarray}
Thus, the interaction modifies the radial behavior of dark energy compared with the non-interacting case. Depending on the values of $w$ and $\epsilon$, the dark-energy contribution can become more important in the central region and influence the final stage of collapse. \\

Using the above solution (\ref{36}) in the dust conservation equation (\ref{32}) gives
\begin{eqnarray} \label{37}
\rho_m' + \frac{2}{\chi}\rho_m = -\epsilon\rho_{\rm de}^0(t)\chi^{\epsilon - 2(1+w)-1}.
\end{eqnarray}
Multiplying by the integrating factor $\chi^2$, we obtain
\begin{eqnarray} \label{38}
(\chi^2\rho_m)' = -\epsilon\rho_{\rm de}^0(t)\chi^{\epsilon - 2w}.
\end{eqnarray}
Integrating with respect to $\chi$, we get
\begin{eqnarray} \label{39}
\chi^2\rho_m = -\frac{\epsilon\rho_{\rm de}^0(t)}{\epsilon - 2w}\chi^{\epsilon - 2w} + \rho_m^0(t),
\end{eqnarray}
where $\rho_m^0(t)$ is another function of integration that depends only on time. This leads to
\begin{eqnarray} \label{40}
\rho_m(t,\chi) = \chi^{-2}\left[\rho_m^0(t) - \frac{\epsilon\rho_{\rm de}^0(t)}{\epsilon - 2w}\chi^{\epsilon - 2w}\right].
\end{eqnarray}
Consequently, the total energy density that determines the geometry is $\rho_t(t,\chi) = \rho_m(t,\chi) + \rho_{\rm de}(t,\chi)$, or explicitly,
\begin{eqnarray} \label{41}
\rho_t(t,\chi) = \chi^{-2}\left[\rho_m^0(t) - \frac{\epsilon\rho_{\rm de}^0(t)}{\epsilon - 2w}\chi^{\epsilon - 2w}\right] + \rho_{\rm de}^0(t)\chi^{\epsilon - 2(1+w)}.
\end{eqnarray}
The pressure (\ref{17}) is generated only by the dark-energy component and is therefore,
\begin{eqnarray} \label{42}
P(t,\chi) = w\rho_{\rm de}^0(t)\chi^{\epsilon - 2(1+w)}.
\end{eqnarray}

\subsection{Mass Function and Horizon Formation}

The total density is related to the generalized mass function through $\rho_t(t,\chi) = 2M'(t,\chi)/\chi^2$. Using the expression (\ref{41}) for $\rho_t$, we obtain
\begin{eqnarray} \label{43}
2M' = \rho_m^0(t) - \frac{\epsilon\rho_{\rm de}^0(t)}{\epsilon - 2w}\chi^{\epsilon - 2w} + \rho_{\rm de}^0(t)\chi^{\epsilon - 2w}.
\end{eqnarray}
This further gives,
\begin{eqnarray} \label{44}
2M' = \rho_m^0(t) - \frac{2w\rho_{\rm de}^0(t)}{\epsilon - 2w}\chi^{\epsilon - 2w}.
\end{eqnarray}
Integrating with respect to $\chi$, the mass function becomes
\begin{eqnarray} \label{45}
M(t,\chi) = M_0(t) + \frac12\rho_m^0(t)\chi - \frac{w\rho_{\rm de}^0(t)}{(\epsilon - 2w)(\epsilon - 2w + 1)}\chi^{\epsilon - 2w + 1}.
\end{eqnarray}
The first term represents the central mass contribution, the second term is the contribution from the dust component, and the last term represents the modification produced by the interacting dark-energy sector. Thus, the interaction changes the effective gravitational mass of the collapsing object and consequently modifies the location and evolution of the apparent horizon. \\

In the limit $\epsilon \to 0$, this reduces to the non-interacting case. As the metric function (\ref{2}) is
\begin{eqnarray*}
f(t,\chi) = 1 - \frac{2M(t,\chi)}{\chi},
\end{eqnarray*}
and thus the apparent horizon is determined by
\begin{eqnarray} \label{46}
f(t,\chi_H) = 0.
\end{eqnarray}

\subsection{Curvature Invariants and Singularity Structure}
The curvature invariants retain the same geometrical form as in the non-interacting case, but their behavior is modified through the new density and mass functions. Therefore, the Kretschmann scalar is
\begin{eqnarray} \label{47}
K = \frac{48M^2}{\chi^6} - \frac{16M}{\chi^3}(2\rho_t - P) + 8\rho_t^2 - 8\rho_t P + 4P^2.
\end{eqnarray}
The constant interaction model shows that energy exchange can modify and, for suitable parameter choices, weaken the central divergence compared with the non-interacting collapse. In particular, for suitable values of $\epsilon$, the divergent part of the density can be cancelled by a fine-tuned relation between the dust and dark-energy coefficients. However, the constant interaction acts uniformly throughout the cloud and does not provide a mechanism that becomes increasingly dominant only in the high-density central region.

\subsection{Regularity Conditions}

Note that if $\epsilon \to 0$ then the solution for the mass function reduces to the non-interacting case. One important remark is that if we take
\begin{eqnarray} \label{48}
\epsilon \geq 2(1+w)
\end{eqnarray}
Under this scenario, the energy density and pressure at the center remain finite, as long as the following condition holds
\begin{eqnarray} \label{49}
\rho_m^0(t) = \frac{\epsilon\rho_{\rm de}^0(t)}{\epsilon - 2w}.
\end{eqnarray}
This condition represents a fine-tuning between the dust and dark-energy components that cancels the divergent $\chi^{-2}$ term in the total density. Under this condition, the leading divergent contribution to the density is removed, and the pressure remains finite provided the same central scaling condition is satisfied. Importantly, the fine tuning condition~(\ref{49}) follows directly from the energy density~(\ref{41}) and is required to avoid divergence as $\chi \to 0$. However, the mass function~(\ref{45}) cannot yield a regular black hole, even with this cancellation, since  $\lim_{\chi \to 0} M(t,\chi) = M_0(t) \neq 0$. This leaves a weak singularity, in contrast to the non-interacting case~(\ref{19}) where a strong singularity is observed \cite{tipler78, brien99, clarke85, Vertogradov_2025}. \\
However, the solution for the mass function does not generally results in a regular black-hole core due to central contribution $M_0(t)$ in the mass function~(\ref{45}). For any fixed $t>0$ with $M_0(t)\neq0$, one has
\begin{eqnarray} \label{50}
\lim_{\chi\to0}M(t,\chi)=M_0(t)\neq0.
\end{eqnarray}
Therefore, the metric function behaves as
\begin{eqnarray} \label{51}
f(t,\chi)=1-\frac{2M_0(t)}{\chi},
\end{eqnarray}
and the central curvature remains singular. In particular, cancellation of the divergent density term does not by itself guarantee a regular geometry. \\ 

At the collapse endpoint $t\to0$, it is possible for the central mass contribution to vanish. For example, with the choice
\begin{eqnarray}  \label{52}
M_0(t)=\mu t,
\end{eqnarray}
we have $M_0(t)\to0$ as $t\to0$. However, this does not by itself establish regularity, since a regular spherical center requires the stronger condition
\begin{eqnarray} \label{53}
M(t,\chi)=\mathcal{O}(\chi^3)
\end{eqnarray}
as $\chi\to0$. Thus, the constant-interaction model may soften the central behavior under suitable fine-tuning, but it does not generically produce a regular black-hole core. \\

To investigate the visibility of the central region, we assume a radial null geodesic given by
\begin{eqnarray} \label{54}
\frac{dt}{d\chi} = \frac{2}{1 - \frac{2M(t,\chi)}{\chi}}.
\end{eqnarray}
If this geodesic terminates at the center in the past and is future-directed, then information from this region can in principle reach an external observer (similar case discussed in Ref \cite{vitali95}). This may happen if the quantity $\chi_0$ defined as
\begin{eqnarray} \label{55}
\lim_{t \to 0, \chi \to 0} \frac{dt}{d\chi} = \chi_0,
\end{eqnarray}
has a positive and finite value. The existence of such a finite $\chi_0$ indicates that outgoing null geodesics can emerge from the center, making the weak singularity potentially globally visible. \\ 

Let us substitute the mass function~(\ref{45}) into the geodesic equation and assume, for simplicity, that
\begin{eqnarray} \label{56}
M_0(t) = \mu t, \quad \mu > 0,
\end{eqnarray}
\begin{eqnarray} \label{57}
\rho_m^0(t) = \gamma t, \quad \gamma > 0,
\end{eqnarray}
\begin{eqnarray} \label{58}
\rho_{\rm de}^0(t) = \delta t^{-\epsilon + 2w}, \quad \delta > 0,
\end{eqnarray}
and take the limit $\chi\to0, t\to0$. We then arrive at the algebraic equation
\begin{eqnarray} \label{59}
\chi_0 = \frac{2}{1 - 2\mu \chi_0 + \dfrac{2w\delta}{(\epsilon - 2w)(\epsilon - 2w + 1)} \chi_0^{-\epsilon + 2w}}
\end{eqnarray}
or equivalently,
\begin{eqnarray} \label{60}
2\mu \chi_0^3 - \chi_0^2 + 2\chi_0 - \frac{2w\delta}{(\epsilon - 2w)(\epsilon - 2w + 1)}\chi_0^{-\epsilon + 2w + 2} = 0.
\end{eqnarray}
Equation~(\ref{60}) can admit a positive finite root in the interval $(0,1)$ for appropriate values of the parameters. In particular, a sufficient parameter condition can be obtained by evaluating the algebraic equation at $\chi_0=1$, which gives
\begin{eqnarray} \label{61}
1 + 2\mu - \frac{2w\delta}{(\epsilon - 2w)(\epsilon - 2w + 1)} > 0.
\end{eqnarray}
The existence of such a positive finite root indicates that outgoing null geodesics can emerge from the center, making the weak singularity potentially globally visible. This represents a possible violation of cosmic censorship.
Another notable aspect is that the solution fails to be \textbf{ asymptotically flat}, and it contains an outer apparent horizon analogous to the cosmological one. \\ 

For $w < 0$:
\begin{itemize}
\item If $\epsilon > 0$: The condition $w < \frac{\epsilon}{2}$ is automatically satisfied for all $w < 0$. No extra restriction.
\item If $\epsilon < 0$: The condition $w > \frac{\epsilon}{2}$ imposes a lower bound on $w$. Since $w < 0$, this means:
\begin{eqnarray} \label{62}
\frac{\epsilon}{2} < w < 0.
\end{eqnarray}
\end{itemize}
Again, densities cannot be constant for any condition except $\rho_m^0(t) = 0$, but this would imply that the cloud is purely composed of dark energy. As a consequence, although the density divergence can be softened through the fine-tuned interaction, the condition required for a completely regular black-hole core cannot generally be satisfied. Therefore, the constant interaction model does not produce a fully regular black hole solution. \\
This motivates considering a more realistic scenario in which the interaction strength depends on the radial position and increases towards the center of the collapsing cloud. Such an interaction can become important precisely in the high-density region where singularity formation takes place.

\section{Radially Varying Interaction Model}
\label{sec:varying}

The previous model assumes that the interaction takes place throughout the collapsing cloud with a constant interaction strength. In this section, we consider a more general situation in which the interaction depends on the radial coordinate and becomes increasingly important toward the central region. The conservation equations for the dust and dark-energy components are written as
\begin{eqnarray}
\frac{\partial \rho_m}{\partial \chi}
+\frac{2}{\chi}\rho_m
&=&
-\frac{\epsilon(\chi)}{\chi}\rho_{\rm de},
\label{var1}
\end{eqnarray}
and
\begin{eqnarray}
\frac{\partial \rho_{\rm de}}{\partial \chi}
+\frac{2(1+w)}{\chi}\rho_{\rm de}
&=&
\frac{\epsilon(\chi)}{\chi}\rho_{\rm de}.
\label{var2}
\end{eqnarray}
Here, $\epsilon(\chi)$ characterizes the radial interaction between the dust and dark-energy sectors. The radial dependence of $\epsilon(\chi)$ allows the interaction coefficient to become significant in the high-density central region while remaining weaker in the outer part of the collapsing configuration. \\

Solving Eq.~(\ref{var2}), we obtain 
\begin{eqnarray} 
\rho_{\rm de}(t,\chi) &=& \rho_{\rm de}^{0}(t) \chi^{-2(1+w)} \exp\left[ \int \frac{\epsilon(\chi)}{\chi}d\chi \right], 
\label{var3} 
\end{eqnarray} 
where $\rho_{\rm de}^{0}(t)$ is an integration function. Substituting Eq.~(\ref{var3}) into Eq.~(\ref{var1}) and using the integrating factor $\chi^2$, the dust density is obtained as 
\begin{eqnarray} 
\rho_m(t,\chi) &=& \frac{1}{\chi^2} \Bigg[ B_0(t)
-
\rho_{\rm de}^{0}(t)
\int
\epsilon(\chi)
\chi^{-2w}
\exp\left(
\int\frac{\epsilon(\chi)}{\chi}d\chi
\right)d\chi
\Bigg],
\label{var4}
\end{eqnarray}
where $B_0(t)$ is a function of integration. \\
The corresponding mass function is obtained from
\begin{eqnarray}
\rho_t= \frac{2M'}{\chi^2},
\end{eqnarray}
and the formal solution is
\begin{eqnarray}
M(t,\chi) = M_0(t) + \frac{1}{2} \int \chi^2 \left(\rho_m+\rho_{\rm de}\right)d\chi,
\label{var5}
\end{eqnarray}
where $M_0(t)$ is an integration function. \\

A regular center requires both the energy density and the mass function to remain finite as $\chi\rightarrow0$. In particular, we require
\begin{itemize}
    \item The densities should be constant (finite) at the center, i.e.,
    $\lim_{\chi\rightarrow0}\rho_m(t,\chi)<\infty$ and $\lim_{\chi\rightarrow0}\rho_{\rm de}(t,\chi)<\infty$,
    Note that this is possible only if the square bracket's expression is considered as zero;
    \item The limit \( \lim_{\chi \to 0} M(t, \chi) = 0 \) should be held.
\end{itemize}
The first condition requires cancellation of any inverse powers of $\chi$ appearing in the density. The second condition guarantees that the mass enclosed by a vanishingly small sphere tends to zero. In the constant-interaction model considered previously, the first condition could be imposed through a suitable relation between the integration constants, but the second condition could not be simultaneously satisfied. We now show that a radially varying interaction can satisfy both requirements.

\subsection{Explicit Example and Regular Solution}
We consider the following radial dependence of the interaction:
\begin{eqnarray}
\epsilon(\chi)
&=&
2(1+w)-a\chi,
\label{var8}
\end{eqnarray}
where $\alpha(t)$ determines the radial variation of the interaction. This form is particularly useful because it naturally leads to finite densities at the center while allowing the magnitude of the interaction to become increasingly important toward the central region. For the phantom regime $w<-1$, $\epsilon(\chi)$ becomes increasingly negative as $\chi\rightarrow0$, so that $|\epsilon(\chi)|$ grows without bound toward the center. \\

Note that with this choice,
\begin{eqnarray}
\frac{\epsilon(\chi,t)}{\chi} = \frac{2(1+w)}{\chi} - \alpha(t).
\end{eqnarray}
Substituting Eq.~(\ref{var8}) into Eq.~(\ref{var3}), we obtain
\begin{eqnarray}
\rho_{\rm de}(t,\chi)
&=&
\rho_{\rm de}^{0}(t)
\chi^{-2(1+w)}
\exp\left[
\int\left(\frac{2(1+w)}{\chi}-\alpha\right)d\chi
\right].
\end{eqnarray}
This simplifies to,
\begin{eqnarray}
\rho_{\rm de}(t,\chi) = \rho_{\rm de}^{0}(t)e^{-\alpha\chi}.
\label{var10}
\end{eqnarray}
Thus, the dark-energy density remains finite at the center,
\begin{eqnarray}
\rho_{\rm de}(t,0) = \rho_{\rm de}^{0}(t).
\label{var11}
\end{eqnarray}
Using Eq.~(\ref{var10}) in Eq.~(\ref{var4}), and noting that 
\begin{eqnarray*}
\int \frac{\epsilon(\chi)}{\chi}d\chi = 2(1+w)\ln\chi - \alpha\chi,
\end{eqnarray*}
we have 
\begin{eqnarray*}
e^{\int \frac{\epsilon(\chi)}{\chi}d\chi} = \chi^{2(1+w)}e^{-\alpha\chi}.
\end{eqnarray*}
 Therefore,
\begin{eqnarray}
\rho_m(t,\chi) &=& \frac{1}{\chi^2} \Bigg[ B_0
-
\rho_{\rm de}^{0}(t)
\int
(2(1+w)-\alpha \chi)
\chi^{2}e^{-\alpha \chi}
d\chi
\Bigg],
\end{eqnarray}
where $B_0$ is constant of integration. After evaluating the integral, we get
\begin{eqnarray}
\rho_m(t,\chi) = \chi^{-2}\left[B_0 - \frac{\rho_{\rm de}^{0}(t)}{\alpha^2}e^{-\alpha\chi}\left(\alpha^2\chi^2 - 2w \alpha\chi - 2w\right)\right].
\label{var12}
\end{eqnarray}
The leading $\chi^{-2}$ divergence is removed by choosing the integration function according to
\begin{eqnarray}
B_0 &=& -\frac{2w}{\alpha^2} \rho_{\rm de}^{0}(t). 
\label{var13} 
\end{eqnarray} 
This relation represents a fine-tuning between the dust and dark-energy integration functions. It precisely removes the leading central divergence of the dust density. Consequently,
\begin{eqnarray} 
\rho_m(t,\chi) = \frac{\rho_{\rm de}^{0}(t)}{\chi^2} \left[ \frac{2w}{\alpha^2} \left(e^{-\alpha\chi}-1\right) + \frac{2w\chi}{\alpha}e^{-\alpha\chi} - \chi^2e^{-\alpha \chi} \right].
\label{var14}
\end{eqnarray}
The dust density around the center is
\begin{eqnarray}
\lim_{\chi\rightarrow0}\rho_m(t,\chi) = \rho_m (t,0)= -(1+w)\rho_{\rm de}^{0}(t).
\label{var20}
\end{eqnarray}
Thus, the dust density is finite at the center, with its value determined entirely by the dark-energy density. For a physical regular center with non-negative dust density, assuming $\rho_{\rm de}^{0}(t)>0$, we require $w\leq -1$. For the phantom regime $w<-1$, the dust density is positive at the center, while the limiting case $w=-1$ gives zero dust density at the center. In the latter case, the central core approaches a de Sitter configuration. \\

\begin{theorem}[Regularity Criterion] \label{regularity}
For the regular black hole solution obtained via the radially varying interaction mechanism, the central dust density is $\rho_m(t,0) = -(1+w)\rho_{\rm de}^{0}(t)$. The requirement $\rho_m(t,0) \geq 0$ implies $w \leq -1$. Thus, physically admissible regular black hole solutions exist only for $w \leq -1$: for $w < -1$ (phantom regime) the dust density is positive, while for $w = -1$ (cosmological constant) the dust density vanishes and the core becomes pure de Sitter.
\end{theorem}

\begin{figure}[h]
\centering
\includegraphics[width=1.0\textwidth]{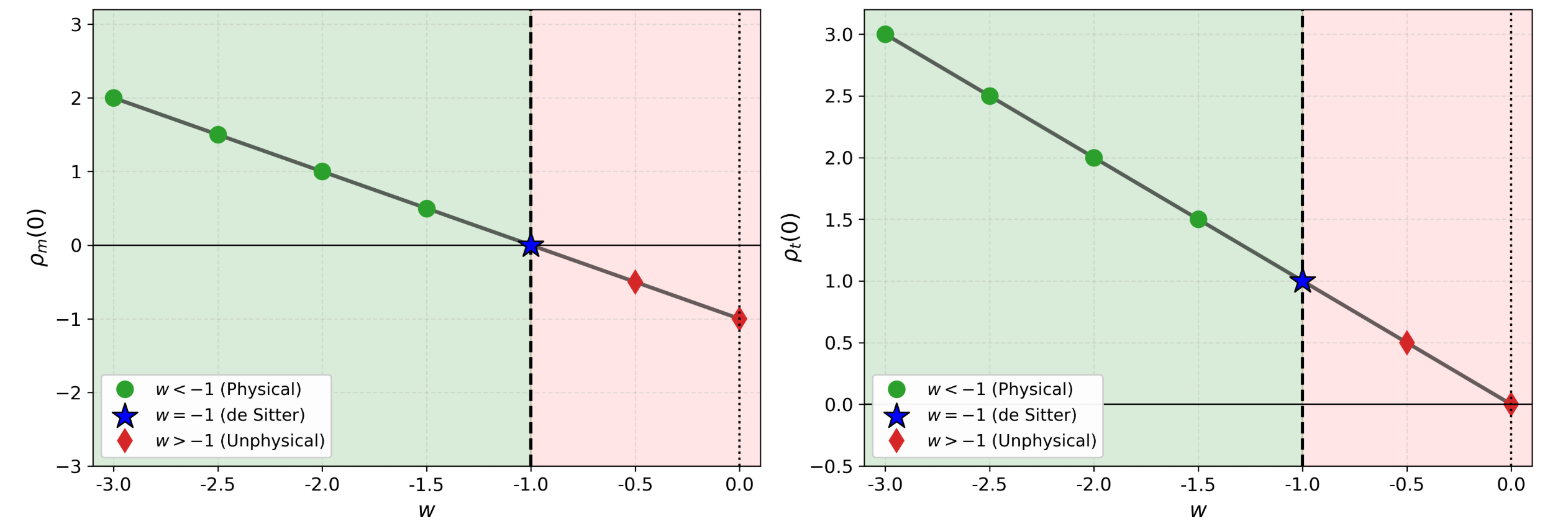}
\caption{(a) Dust density  $\rho_m (0)$, and (b) the total density $\rho_t(0)$ at the center ($\chi=0$) as a function of $w$. The total density remains positive for all $w$. While the dust density alone vanishes for $w=-1$ (de Sitter core), remains positive for $w<-1$ (physical) and becomes negative for $w>-1$ (unphysical). }
\label{fig:density}
\end{figure}

Similarly, the total density is 
\begin{eqnarray}
\rho_t(t,\chi) = \frac{\rho_{\rm de}^{0}(t)}{\chi^2} \left[ \frac{2w}{\alpha^2} \left(e^{-\alpha\chi}-1\right) + \frac{2w\chi}{\alpha}e^{-\alpha\chi} \right].
\label{var16}
\end{eqnarray}
therefore, the total density around the center is given by
\begin{eqnarray}
\lim_{\chi\rightarrow0}\rho_t(t,\chi) = \rho_t(t,0)= -w\rho_{\rm de}^{0}(t).
\label{var18}
\end{eqnarray}
Using $2M' = \chi^2\rho_t$, we have
\begin{eqnarray}
M' = \frac{w\rho_{\rm de}^{0}(t)}{\alpha^2}(e^{-\alpha\chi} - 1) + \frac{w\rho_{\rm de}^{0}(t)}{\alpha}\chi e^{-\alpha\chi}.
\end{eqnarray}
Integrating,
\begin{eqnarray}
M(t,\chi) = M_0(t) - \frac{w\rho_{\rm de}^{0}(t)}{\alpha ^2}\chi - \frac{w\rho_{\rm de}^{0}(t)}{\alpha^3}e^{-\alpha\chi}(\alpha\chi + 2).
\label{var23}
\end{eqnarray}
The remaining integration freedom in the mass function is fixed by the regularity condition. Since the mass enclosed within a vanishingly small sphere must vanish, we require
\begin{eqnarray}
M_0(t) = \frac{2w\rho_{\rm de}^{0}(t)}{\alpha^3(t)}.
\label{var21}
\end{eqnarray}
With this choice, the mass enclosed within a sphere of vanishing radius tends to zero. The leading $\chi^3$ behavior is the standard behavior required for a regular spherical center. \\

Finally, we obtain the regular black hole solution,
\begin{align}
M(t, \chi) = M_0 \left[ 1 - \frac{\alpha \chi}{2} - \left( 1 + \frac{\alpha \chi}{2} \right) e^{-\alpha \chi} \right]
\end{align}
Similarlyy, the metric function is given by the full expression
\begin{eqnarray}
f(t,\chi) = 1 - \frac{4w}{\alpha^3\chi}\rho_{\rm de}^{0}(t)\left[1 - \frac{\alpha\chi}{2} - \left(1 + \frac{\alpha\chi}{2}\right)e^{-\alpha\chi}\right].
\label{metric_full}
\end{eqnarray}
Near the center, this reduces to
\begin{eqnarray}
f(t,0)=1, \qquad f'(t,0)=0,
\label{var27}
\end{eqnarray}
and the metric function has a regular Taylor expansion at the center. \\
The Kretschmann scalar $K$ of our solution at the center is
\begin{eqnarray}
K(t,0)= \frac{2\alpha^6M_0^{\ 2}(t)}{3},
\label{var28}
\end{eqnarray}
which is finite, confirming the regularity of the solution.
\begin{figure}[h]
\centering
\includegraphics[width=0.75\textwidth]{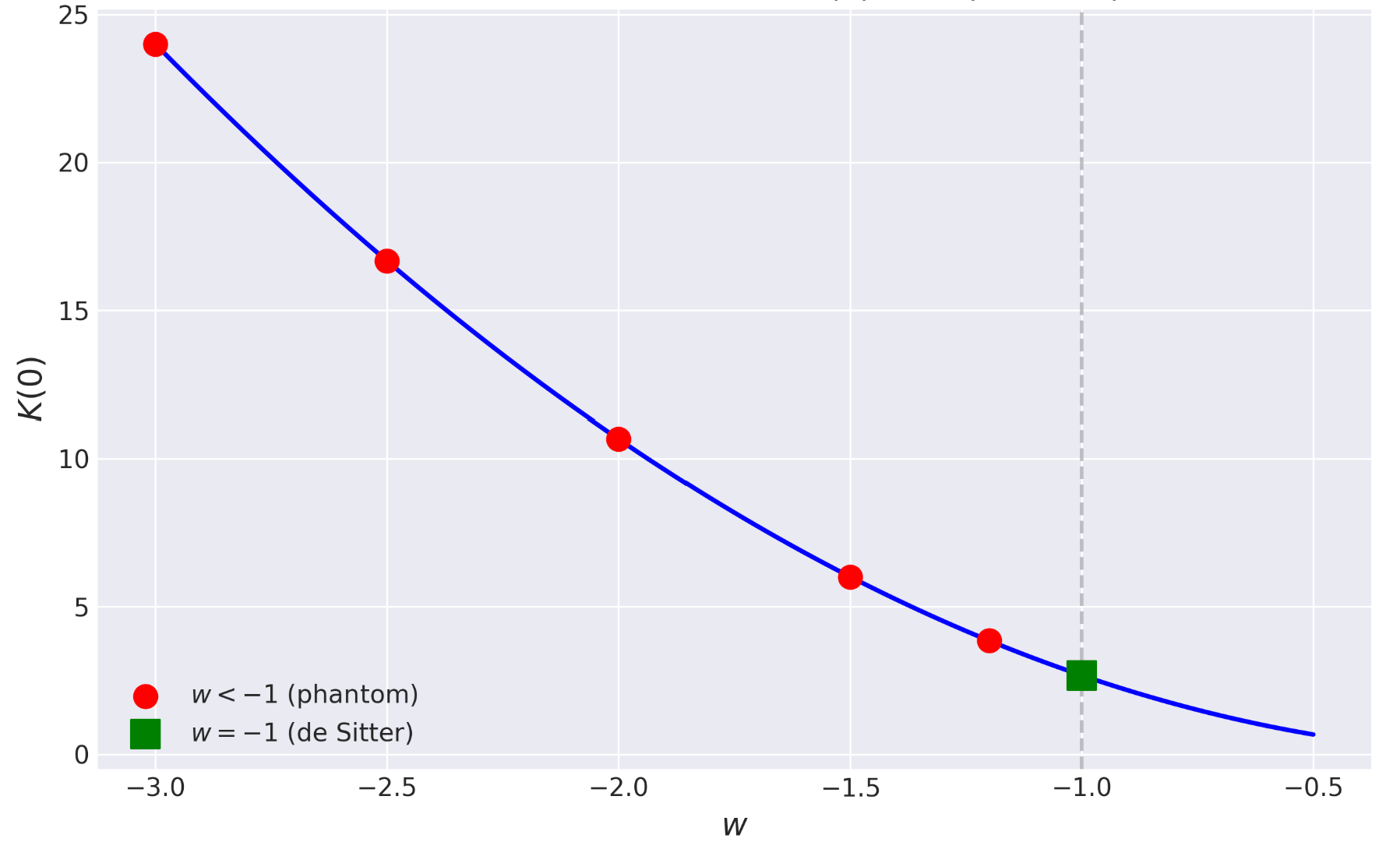}
\caption{The Kretschmann scalar $K$ at the center ($\chi=0$) as a function of $w$. It remains finite for all $w$. }
\label{fig:curvature}
\end{figure} 
The regularity conditions therefore lead to a finite total density and a mass function behaving as $M\sim\chi^3$. In particular, the full expressions for the densities at the center are
\begin{eqnarray} 
\rho_{\rm de}(t,0) &=& \rho_{\rm de}^{0}(t),\\
\rho_m(t,0) &=& -(1+w)\rho_{\rm de}^{0}(t),\\
\rho_t(t,0) &=& -w\rho_{\rm de}^{0}(t),
\end{eqnarray}
while the mass function, metric function and behave as
\begin{eqnarray}
M(t,\chi) &\sim& -\frac{w\rho_{\rm de}^{0}(t)}{6}\chi^3,\\
f(t,\chi) &\sim& 1 + \frac{w\rho_{\rm de}^{0}(t)}{3}\chi^2, \\
\lim_{\chi\rightarrow0}K &<& \infty.
\label{var29}
\end{eqnarray}
For $-1<w<0$, the central dust density is negative and is therefore non-physical if $\rho_{\rm de}^{0}(t)>0$. For $w\leq-1$, the central densities are non-negative. In particular, in the phantom regime $w<-1$, the central dust density is positive, giving a de Sitter-like core with both dust and dark energy present. The metric near the center takes the de Sitter form $f(t,\chi) = 1 - H^2\chi^2$ with $H^2 = -w\rho_{\rm de}^{0}(t)/3 > 0$ for any $w<0$. The limiting case $w=-1$ corresponds to a pure de Sitter core with vanishing central dust density. Moreover, for $\rho_{\rm de}^{0}(t)>0$, a positive mass scale $M_0(t)>0$ requires $\alpha<0$ in the phantom regime. Thus, the interaction mechanism provides a family of regular black hole solutions for $w \leq -1$, where the regularization arises from the energy exchange between dust and dark energy, not from an electric charge as in conventional regular black holes. 
The resulting solution provides an explicit example in which a radially varying interaction removes the central density divergence while maintaining a regular near-center metric. The essential mechanism is the cancellation of the leading $\chi^{-2}$ contribution through the relation in Eq.~(\ref{var13}), followed by the condition in Eq.~(\ref{var21}), which fixes the remaining integration freedom in the mass function. The resulting mass function satisfies $M(t,\chi)\rightarrow0$ as $\chi\rightarrow0$, with the leading behavior $M\sim\chi^3$. \\

Thus, the radially varying interaction provides an analytically tractable mechanism for avoiding the central singularity and generating a regular core. In particular, the interaction becomes increasingly important in magnitude toward the center, while the energy densities remain finite and the enclosed mass vanishes as $\chi^3$. This provides an explicit realization of singularity avoidance through a spatially varying interaction between the dust and dark-energy sectors. 

\subsection{Energy Conditions}

We now examine the viability of the regular black hole solution obtained in the varying interaction model by analyzing the standard energy conditions. For a fluid with total energy density $\rho_t$ and isotropic pressure $P$, the energy conditions are defined as follows: the null energy condition (NEC) requires $\rho_t + P \geq 0$; the weak energy condition (WEC) requires $\rho_t \geq 0$ and $\rho_t + P \geq 0$; the strong energy condition (SEC) requires $\rho_t + P \geq 0$ and $\rho_t + 3P \geq 0$; and the dominant energy condition (DEC) requires $\rho_t \geq 0$ and $\rho_t \geq |P|$. \\

At the center of the regular core, the total energy density and pressure are given by
\begin{eqnarray}
\rho_t(t,0) &=& -w\rho_{\rm de}^{0}(t), \label{ec1}\\
P(t,0) &=& w\rho_{\rm de}^{0}(t). \label{ec2}
\end{eqnarray}
These expressions follow directly from Eqs.~(\ref{var18}) and the equation of state $P = w\rho_{\rm de}$. \\

For the NEC, we obtain
\begin{eqnarray}
\rho_t(t,0) + P(t,0) = -w\rho_{\rm de}^{0}(t) + w\rho_{\rm de}^{0}(t) = 0,
\label{ec3}
\end{eqnarray}
so the NEC is marginally satisfied at the center. This saturation is a characteristic feature of regular black hole cores and is consistent with the presence of a de Sitter-like geometry. \\

The WEC requires $\rho_t \geq 0$. From Eq.~(\ref{ec1}), and assuming $\rho_{\rm de}^{0}(t) > 0$, this imposes
\begin{eqnarray}
w \leq 0.
\label{ec4}
\end{eqnarray}
Together with the requirement of non-negative central dust density, which gave $w \leq -1$ from Eq.~(\ref{var20}), the WEC is satisfied for the phantom regime $w < -1$ and marginally for $w = -1$. \\

The SEC requires $\rho_t + 3P \geq 0$. Substituting Eqs.~(\ref{ec1}) and (\ref{ec2}), we find
\begin{eqnarray}
\rho_t(t,0) + 3P(t,0) = -w\rho_{\rm de}^{0}(t) + 3w\rho_{\rm de}^{0}(t) = 2w\rho_{\rm de}^{0}(t).
\label{ec5}
\end{eqnarray}
For $w < 0$, which includes the phantom regime $w < -1$, this quantity is strictly negative. Hence, the SEC is violated. This violation is not only permissible but, in fact, necessary for the formation of a regular core, as it provides the repulsive gravitational effect required to halt the collapse and prevent the formation of a curvature singularity. This is consistent with the behavior of known regular black hole solutions, including the Bardeen and Hayward models, where the SEC is also violated in the core region. \\

Finally, the DEC requires $\rho_t \geq |P|$. Using Eqs.~(\ref{ec1}) and (\ref{ec2}), we obtain
\begin{eqnarray}
\rho_t(t,0) - |P(t,0)| = -w\rho_{\rm de}^{0}(t) - |w|\rho_{\rm de}^{0}(t).
\label{ec6}
\end{eqnarray}
For $w < 0$, we have $|w| = -w$, so
\begin{eqnarray}
\rho_t(t,0) - |P(t,0)| = -w\rho_{\rm de}^{0}(t) + w\rho_{\rm de}^{0}(t) = 0.
\label{ec7}
\end{eqnarray}
Thus, the DEC is marginally satisfied at the center. \\

We note that the violation of the SEC is a generic feature of regular black hole spacetimes with de Sitter-like cores. In particular, it is this violation that enables the transition from the collapsing phase to a regular core, effectively avoiding the singularity. The present solution therefore satisfies all physically reasonable energy conditions except the SEC, which is expected for a configuration dominated by dark energy. \\

It is important note that, for $-1<w<0$, the total fluid satisfies energy conditions, but the dust component alone does not, as it  acquires a negative energy density in this regime, making the configuration unphysical for ordinary dust. If one were to interpret this negative density as exotic matter, the weak energy condition could be restored, but such an interpretation would require a physical mechanism for generating exotic fluid. In this work , we therefore restrict our analysis to the physically consistent regime $w\leq -1$, where both the total fluid and the dust component satisfy the energy conditions.

\section{Discussion}
It is instructive to compare our interaction-based regularization mechanism with the well-known charge-based regular black holes supported by nonlinear electrodynamics. In those models, the electric charge $g$ serves as the regularization parameter, and the Kretschmann scalar at the center behaves as $1/g^6$. For $g \neq 0$, the solution is regular, but as $g \to 0$, the singularity reappears. These models face a significant physical limitation: astrophysical black holes are expected to be neutral, and any charge would evaporate over time through accretion or the Penrose process, inevitably leading to singularity formation. \\

In our model, the interaction parameter $\alpha$ plays a role similar to the charge $g$, but with an important difference. For $\alpha \neq 0$, the varying interaction yields a regular black hole with a finite Kretschmann scalar at the center [see Eq.~(\ref{var28})],
\begin{equation}
\lim_{\chi\to0} K < \infty.
\end{equation}
However, if $\alpha \to 0$, the interaction becomes constant and the solution reduces to the singular constant-interaction model [see Eq.~(\ref{45})]. Thus, like the charge $g$, the parameter $\alpha(t)$ must remain non-zero to maintain regularity. The crucial distinction is that $\alpha$ does not evaporate. During the collapse, the interaction should grow toward the center. A decreasing $\alpha(t)$ ensures that the interaction between dust and dark energy becomes stronger as the collapse proceeds. Thus, $\alpha(t)$ must monotonically decrease ($\dot \alpha(t) < 0$) during collapse, but it stabilizes at a constant, non-zero value at the final stage of collapse. This ensures that the black hole remains regular forever, unlike charge-based models where evaporation inevitably leads to singularity formation. \\
It is important to emphasize that our regular solution is valid only in the inner region. The interaction between dust and dark energy prevents singularity formation at the center, but the solution must be extended to larger radii by matching it to an exterior geometry at some junction surface $\chi = R$. This matching can be performed using the standard Israel junction conditions, requiring continuity of the metric, extrinsic curvature, and matter fields across the junction. To join the interior regular core [Eqs.~(\ref{var23}) and (\ref{var21})] with an exterior geometry at the junction surface $\chi = R$, one must impose continuity of the pressures, energy densities, and mass functions across the junction. \\

Following the matching procedure developed in Ref.~\cite{Vertogradov_2025} for the radiation case, we adapt the junction conditions for our dark energy model with general $w$. The matching conditions require that at $\chi = R$, the pressures, energy densities, and mass functions of both solutions coincide:
\begin{eqnarray}
P_{\rm int}(R) = P_{\rm ext}(R), \qquad
\rho_{\rm int}(R) = \rho_{\rm ext}(R), \qquad
M_{\rm int}(R) = M_{\rm ext}(R).
\end{eqnarray}
These three conditions can always be satisfied because the exterior solution contains three arbitrary functions of time: $M_0^{\rm ext}(t)$, $\rho_m^0(t)$, and $\rho_{\rm de}^0(t)$. This freedom allows for a smooth transition between the interior regular core and the exterior spacetime. \\

To proceed with the matching, we introduce an auxiliary function $A(t)$ defined as
\begin{eqnarray}
A(t) \equiv \frac{\rho_{\rm de}^0(t)}{3\alpha^3}.
\end{eqnarray}
Using this definition, the pressure matching at $\chi = R$ can be achieved if the exterior dark-energy function satisfies
\begin{eqnarray}
\rho_{\rm de}^0(t) = \frac{3A\alpha ^2}{2} R^{-2w} e^{-\alpha R}.
\end{eqnarray}
Having matched the pressures, we next match the energy densities at $\chi = R$. This requires the exterior dust function to satisfy
\begin{eqnarray}
\rho_m^0(t) = -\frac{A}{2} + \frac{A}{2}\left(1 + \alpha R\right)e^{-\alpha R}.
\end{eqnarray}
Finally, matching the mass functions at $\chi = R$ determines the exterior central mass parameter through
\begin{eqnarray}
M_0^{\rm ext}(t) = M_0(t) + \frac{A R}{4}\left[3 - \left(\alpha R - 2\right)e^{-\alpha R}\right] + \frac{A \alpha ^2}{2}R^2 e^{-\alpha R}.
\end{eqnarray}
If these three conditions are satisfied at the junction radius $\chi = R$, the interior and exterior solutions can be smoothly joined. Nevertheless, these expressions are somewhat complicated, and it is worth recalling that the interaction function $\epsilon(\chi)$ [Eq.~(\ref{var8})] was chosen as a phenomenological example. Other functional forms that grow sufficiently rapidly toward the center could also produce regular solutions, suggesting that the regularization mechanism is robust. \\
For $w < -1$, the central dust density is positive [Eq.~(\ref{var20})] and the solution is physically acceptable. For $-1 < w < 0$, the central dust density becomes negative, making the solution unphysical. The case $w = -1$ corresponds to a pure de Sitter core. Thus, our model naturally favors the phantom regime for realistic regular black holes. \\

Unlike charge-based models where the regularization parameter can evaporate, our interaction-based regularization is stable. The parameter $\alpha(t)$ takes a constant, non-zero value after collapse, ensuring the black hole remains singularity-free indefinitely. This provides a physically motivated alternative to charge-based regular black holes, where the regularization arises from the energy exchange between dust and dark energy rather than from an electric charge that can evaporate.

\section{Conclusion}

The resolution of the central singularity in gravitational collapse remains one of the most profound open questions in gravitational physics. While regular black hole models typically rely on exotic matter or nonlinear electrodynamics, the physical origin of the regularization mechanism often remains obscure. In this work, we have demonstrated that a radially varying interaction between dust and dark energy provides a natural and physically motivated mechanism for singularity avoidance, grounded in the modern cosmological understanding of dark energy. \\
We have systematically analyzed three scenarios. (i) In the non-interacting case, the dust and dark-energy sectors evolve independently, leading to a divergent dust density~(\ref{14}) $\rho_m \sim \chi^{-2}$, a mass function ~(\ref{19}) with a linear term $M \sim \chi$, and a Kretschmann scalar~(\ref{30}) that diverges ($\lim_{\chi \to 0} K \to \infty$) as $\chi \to 0$. (ii) Introducing a constant interaction softens the density divergence through fine-tuning [Eq.~(\ref{49})], but the mass function~(\ref{50}) retains a non-zero central contribution $M_0(t)$, preventing the formation of a fully regular core. (iii) Only when the interaction acquires a radial dependence, becoming increasingly important toward the center, do both the density divergence and the central mass contribution vanish simultaneously~[(\ref{var13}) and (\ref{var21})]. The specific choice $\epsilon(\chi) = 2(1+w)/\chi - \alpha(t)$ [Eq.~(\ref{var8})] yields an explicit regular solution~(\ref{var29}) with $M \sim \chi^3$, a finite Kretschmann scalar~(\ref{var28}) $\lim_{\chi \to 0} K < \infty$, and a de Sitter-like metric~(\ref{var27}) near the center. Furthermore, as established in Theorem~\ref{regularity}, the physical admissibility of these solutions requires $w \leq -1$, ruling out the quintessence-like regime $w > -1$ where the central dust density becomes negative. \\
A key strength of our model is its generality. The regularization mechanism works for the entire phantom branch $w \leq -1$, with the cosmological constant case $w = -1$ corresponding to a pure de Sitter core and the phantom regime $w < -1$ yielding a de Sitter-like core with positive central dust density [Eq.~(\ref{var20})]. The energy conditions are satisfied~(\ref{ec3})--(\ref{ec7}), with the NEC, WEC, and DEC holding, while the SEC is violated Eq.~(\ref{ec5}) — a necessary feature for any regular black hole core. This violation provides the repulsive gravity required to halt collapse and prevent singularity formation. Our model offers a significant advantage over charge-based regular black holes, such as those in nonlinear electrodynamics. In those models, the electric charge $g$ serves as the regularization parameter, but charge can evaporate over time, inevitably leading to singularity formation. In our model, the interaction parameter $\alpha(t)$ stabilizes at a non-zero value after collapse~(\ref{var21})], ensuring the black hole remains regular indefinitely. Moreover, unlike models that postulate exotic fields with no known physical origin, our regularization arises naturally from the energy exchange between two well-established components of the universe: dust and dark energy. During gravitational collapse, as the density increases toward the center, the interaction becomes increasingly significant, effectively transforming the collapsing matter into a regular core without the need for ad hoc exotic matter. \\
The regular solutions~(\ref{var23}) \& (\ref{metric_full}) obtained here describes only the inner region of the black hole. We have provided the necessary junction conditions to match the interior regular core to an exterior geometry at some junction surface $\chi = R$, allowing the construction of a complete spacetime. The exterior solution can then be used for observational studies such as black hole shadows. However, the shape of the shadow alone does not provide direct information about the regular center, because an exterior solution lacking a phase transition toward the center would result in a singular black hole. Still, phase transitions may take place prior to the formation of apparent horizon, accompanied by a strong energy flux that influences the observed specific intensity of the shadow. During gravitational collapse, before the horizon forms, this region could in principle be observed by a distant observer \cite{Vertogradov_2025}. \\

The interaction function~(\ref{var8}) $\epsilon(\chi) = 2(1+w)/\chi - \alpha(t)$ was chosen as a concrete example that admits an analytical solution. Other functional forms that grow sufficiently rapidly toward the center are expected to yield similar regularization effects, suggesting that the mechanism is robust. Future work should explore more general equations of state, including barotropic and polytropic models, and investigate the observational signatures of such phase transitions and energy fluxes.

\backmatter
\bibliographystyle{unsrt}
\bibliography{bibliography}

\end{document}